\documentclass{aastex62}

\usepackage{verbatim} 
\graphicspath{{./}{figures/}}

\received{May 6, 2021}
\accepted{June 14, 2021}
\submitjournal{ApJ}

\shorttitle{Mapping the hidden magnetic field of the quiet Sun}
\shortauthors{Trelles Arjona et al.}

\begin{document}

\title{Mapping the hidden magnetic field of the quiet Sun}

\correspondingauthor{J. C. Trelles Arjona}
\email{jtrelles@iac.es, marian@iac.es, brc@iac.es}

\author[0000-0002-0786-7307]{J. C. Trelles Arjona}
\affiliation{Instituto de Astrof\'isica de Canarias (IAC),V\'ia L\'actea s/n, 38205 San Crist\'obal de La Laguna, Tenerife, Spain}
\affiliation{Dept. Astrof\'isica, Universidad de La Laguna, 38205 San Crist\'obal de La Laguna, Tenerife, Spain}

\author{M. J. Mart\'inez Gonz\'alez}
\affiliation{Instituto de Astrof\'isica de Canarias (IAC),V\'ia L\'actea s/n, 38205 San Crist\'obal de La Laguna, Tenerife, Spain}
\affiliation{Dept. Astrof\'isica, Universidad de La Laguna, 38205 San Crist\'obal de La Laguna, Tenerife, Spain}

\author{B. Ruiz Cobo}
\affiliation{Instituto de Astrof\'isica de Canarias (IAC),V\'ia L\'actea s/n, 38205 San Crist\'obal de La Laguna, Tenerife, Spain}
\affiliation{Dept. Astrof\'isica, Universidad de La Laguna, 38205 San Crist\'obal de La Laguna, Tenerife, Spain}



\begin{abstract}

The Sun is the only star where we can resolve the intricate magnetism that all convective stars harbor. Yet, more than 99\% of its visible surface along the solar cycle (the so-called quiet Sun) is filled with a tangled, unresolved magnetism. These "hidden" fields are thought to store enough magnetic energy to play a role in the heating of the Sun's outer atmosphere, but its field strength is still not constrained. Previous investigations based on the Hanle effect in atomic lines claim a strong magnetization of about 100 G, while the same effect in molecules show a factor of 10 weaker fields. The discrepancy disappears if the magnetic field strength of the hidden is not homogeneous in the solar surface. In this letter, we prove using magnetohydrodynamical simulations that it is possible to infer the average field strength of the hidden quiet Sun magnetic fields using multi-line inversions of intensity profiles in the Zeeman regime. Using this technique with 15 spectral lines in the 1.5 $\mu$m spectral range, we reveal that the spatial distribution of the hidden field is strongly correlated with convection motions, and that the average magnetization is about 46 G. Reconciling our findings with the Hanle ones is not obvious and will require future work on both sides, since it implies an increase of the field strength with height, something that is physically questionable.

\end{abstract}



\section{Introduction} \label{sec:introduction}

The Sun is a magnetized star, sometimes showing so spectacular activity that impacts the Earth and yet, at the same time, having the weakest, smallest-scale, stochastic magnetism that will be never observed in a star. The Sun’s global dynamo produces magnetic fields on large scales, such as sunspots and plages, and turbulent shredding ensures that the magnetic scale spectrum extends over many orders of magnitude down to the magnetic diffusion limit near 10-100 m, far below the present and near-future resolution of solar observations. This so-called “hidden” magnetism \citep{stenflo82} fills the quiet solar atmosphere, those areas outside active regions that cover most of the solar surface at any time during the activity cycle. 

During periods of minimum activity, there are no active regions in the surface and all is quiet. Still, the temperature rises in the chromosphere, and the corona maintains its million-degree plasma, implying that the quiet magnetism can be a principal actor in the heating of the Sun’s outer atmosphere \citep{bellot3}. But, after more than 30 years of studies, the actual role of the quiet magnetism in chromospheric and coronal heating is still an open question. Some studies based on Hanle measurements in atomic lines, point that the hidden magnetism in quiet regions stores enough magnetic energy to potentially heat the chromosphere and corona above quiet regions \citep{trujillo04}. Though, the same method applied to molecular lines brings up much less energy \citep{andres05,kleint11}. These results can be reconciled if the hidden magnetic field of the quiet Sun is much stronger in intergranular lanes than in granules, since the radiation emitted by molecules comes mainly from granules \citep{trujillo04,trujillo03}. There is evidence that this is indeed the reality, but detecting Hanle signatures with some spatial resolution is still an observational challenge \citep{zeuner20,zeuner18,delpino18,dhara19}. So far, the spatial distribution of the hidden field has not been observed nor its strength firmly inferred.

 For weak magnetic fields, like in the Sun’s quiet surface, the broadening induced by Zeeman splitting in intensity can be caused, alternatively, by several mechanisms that compete with magnetic fields: velocity gradients, macro- and micro-turbulence, collisional broadening, temperature, Stark effect, etc. However, the ensuing polarimetric signature of the split Zeeman components is unmistakable. Consequently, studying the polarized spectrum of the Sun’s light seems the best option to get an accurate measurement of solar magnetic fields. But the problem with polarization is that, unlike intensity, the Stokes parameters are signed; they can be either positive or negative, depending on the field geometry and hence, the joint contribution of several elements when spatial resolution is not sufficiently high to resolve such small scales can lead to cancellations, inevitably losing information. This is critical in quiet regions, where the hidden field is tangled below the best spatial resolution achieved with the best solar instrumentation (0.2-0.5”). 

The Stokes I parameter encodes the information of the average magnetic field strength of the hidden fields in quiet regions \citep{stenflo77,andres14}. To alleviate the degeneration between the small broadening induced by the Zeeman effect and the one induced by other sources, we must rely on multiline analysis, since all broadening mechanisms depend on atomic parameters in a different manner than the Zeeman splitting
\citep[see e.g., the increase in robustness of multiline Stokes inversions by][]{riethmuller19}. Using the solar optical atlas from \citet{delbouille73} and measuring the width of many lines, a mean value of 140 G was estimated for the magnetic field in quiet areas \citep{stenflo77}. One step further was done using multiline Bayesian analysis of the Zeeman broadening in an optical \citep{wallace98} and an infrared atlas \citep{wallace03}. It was found that, with 90\% probability, the mean magnetic field is smaller than 186 G and 160 G for optical and infrared data, respectively \citep{andres14}. 

We have used near-infrared data with high spatial resolution along with LTE (Local Thermodynamic Equilibrium) inversions to map the spatial distribution of the hidden magnetic field of the quiet Sun from the intensity of many spectral lines. We, first, have proved the capacities of the multiline inversions using magnetohydrodynamical simulations. And second, we have applied the multiline inversions to real observations.

\section{Observations} 

On August 29, we observed a quiet Sun region using the GRIS instrument \citep{tip2,collados12} attached at the GREGOR telescope \citep{schmidt12} at the Observatorio del Teide. We recorded the four Stokes parameters in a large spectral range (40\,\AA\  so that we have access to many spectral lines) around 1.56 $\mu$m that contains the most sensitive line (at 15648\,\AA) to the Zeeman effect in the optical and near infrared spectrum. We scanned a 62” x 54’’ area, along the scan and slit directions, with a step size of 0.135’’. The integration time per slit position was about 2 s which, together with the good seeing conditions and the help of the adaptive optics system, allowed a signal to noise level of 1600 in polarization and 1000 for the intensity. The adaptive optics system \citep{berkefeld} was successfully locked on granulation during the whole scan, granting a spatial resolution of 0.5’’. The spectral sampling was 40 m\,\AA.  
Standard reduction of the data (bias subtraction, flatfield correction, bad pixels and instrumental cross-talk removal, and demodulation) was performed with dedicated software \citep{Schlichenmaier02}. Further corrections were applied to the data: removal of wavelength-independent stray light, of residual cross-talk from Stokes I to Stokes Q, U and V, and of polarized interference fringes. Finally, we decreased the uncorrelated noise level with a procedure based on Principal Component Analysis \citep{loeve55,rees03} (and see \citet{marian2008a} for the actual application to spectro-polarimetric data). More details on the data and data reduction can be found in \citet{trelles21} and references therein.

\section{Testing Stokes I multiline inversions with numerical simulations} \label{sec:simulations}

To prove that the average strength of an unresolved field can be reliably retrieved from the inversion of only the intensity of many spectral lines (see Table 1 in \citet{trelles21} for the description of the spectral lines used in this work), we have synthesized and inverted theoretical GRIS spectra using the Stokes Inversion based on Response functions inversion code \citep[SIR,][]{SIR} in a MANCHARAY magnetohydrodynamical simulation \citep{khomenko17}. The quiet Sun simulation used in this work was done with the MANCHA3D code \citep{khomenko06,tobias10}. This code solves the equation of non-ideal magnetohydrodynamics, together with a realistic equation of state and non-gray radiative transfer \citep{khomenko12}. The simulation includes the Bierman battery term that provides a seed magnetic field (magnetic fields are generated by local imbalances in electron pressure) that is afterwards amplified by the dynamo action \citep{khomenko17}. In the simulation, the mean value of the magnetic field at  $\tau_{500}$ = 1.0 is 90 G, very similar strength to those derived from solar observations of the quiet Sun (details of the numerical set up of the simulation can be found in \citet{khomenko17}).

For a long time, magnetic field measurements in the quiet Sun using the spectral range around 1.56 $\mu$m have been carried out through the inversion of two neutral Fe lines at 15648\,\AA\  and 15652\,\AA\  which are highly sensitive to magnetic fields \citep{bellot3}. To use all lines available, we first had to infer from the observations the atomic parameters of many lines that were unknown. The atomic parameters can be retrieved from observations if the thermodynamical and magnetic properties of the atmosphere are well known a priori. However, this is never the case, and to obtain the model atmosphere, we need the atomic parameters. We proposed a novel, iterative methodology to compute both the atomic parameters of the lines in our spectral window and the model atmosphere, simultaneously \citep{trelles21}. Once the atomic parameters of all lines are fixed, we use SIR to infer the physical properties of the quiet solar atmosphere from the information contained in the data. SIR solves the radiative transfer equation (RTE) for polarized light assuming Local Thermodynamical Equilibrium and the Zeeman effect, which are very suitable approximations for most photospheric lines, and the 1.5 $\mu$m  lines in particular. SIR, as most inversion codes, searches for the synthetic profiles (obtained solving the RTE and a parametrized model atmosphere) that better match the observed ones. This search is performed by minimizing the quadratic distance (the  merit function) between the synthetic and observed profiles. Given the non-linearity of the problem, the minimization requires an iteration scheme (Levenberg-Marquard in the case of SIR).

In order to better fit the profiles, we allow for variations with depth in the velocity, temperature, and magnetic field strength. To infer the stratification of the atmosphere, SIR uses the so-called nodes, that are particular points along the optical depth where perturbations to the parameters are allowed. In between the nodes, SIR interpolates either with low order polynomials or with splines. The interpolation is regularized, and only smooth solutions are permitted. The complexity of the gradients (in SIR, the number of nodes) is automatically chosen in each pixel by evaluating the amount of information contained in the Stokes profiles (see more details in \citet{deltoro16}). The atmospheric gradients in the simulations are very complex, and the SIR code finds a gradient which differs, in a particular log($\tau_{500}$), from that of the simulation but that reproduces the profiles, i.e., it is another solution to the inverse problem. The global behavior of the gradients, however, is captured by the SIR code. Therefore, from now on, we will work with quantities averaged along the spectral lines formation region, from log($\tau_{500}$) = 0.0 to log($\tau_{500}$) =  -1.2 as deduced from response functions.

\subsection{Full resolution, noiseless case}

We first check the reliability of the inferred magnetic field strength from the inversion of the original simulations, i.e., the ideal case of the original spatial resolution, noise-free simulations. To perform the inversion of the profiles from these ideal simulations, we assume that the microturbulence parameter, that usually accounts for unresolved velocities along the line of sight, is zero. Figure 1 shows the inferred magnetic field strength (averaged along the formation region) from the theoretical spectra in this ideal case. The accuracy of the result is astonishing, and not only the field strength is globally recovered but also many tiny details. The match is even more extraordinary when looking at the scatter plot of Fig. 2 (left panel). The magnetic field strength is very well recovered even down to the smallest values. 

\begin{figure}[ht!]
\plotone{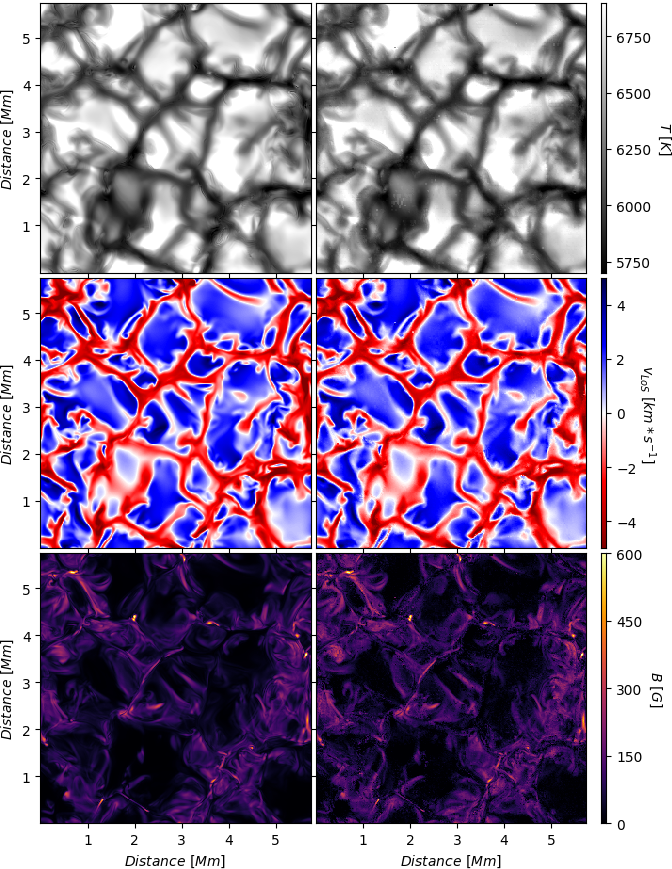}
\caption{Comparison between the simulation (left column) and the results achieved by the inversion (right column). Temperature (at log($\tau_{500}$) = 0.0), line of sight velocity (at log($\tau_{500}$) = -0.2) and magnetic field strength averaged over the formation region of the spectral lines (from log($\tau_{500}$) = 0 to log($\tau_{500}$) = -1.2 as deduced from response functions) in top, middle and bottom row, respectively. The symbol $\tau_{500}$ denotes the optical depth at 500 nm, which is the integral of the product of the opacity at this wavelength and the geometrical length in a photon path. Therefore, it is not representative of a constant geometrical height; more opaque wavelengths tracing, in general, higher layers in the atmosphere. Log($\tau_{500}$) = 0.0 is where continuum at 500 nm is formed, and more negative numbers mean higher in the atmosphere.\label{fig:one}}
\end{figure}

\begin{figure}[ht!]
\plotone{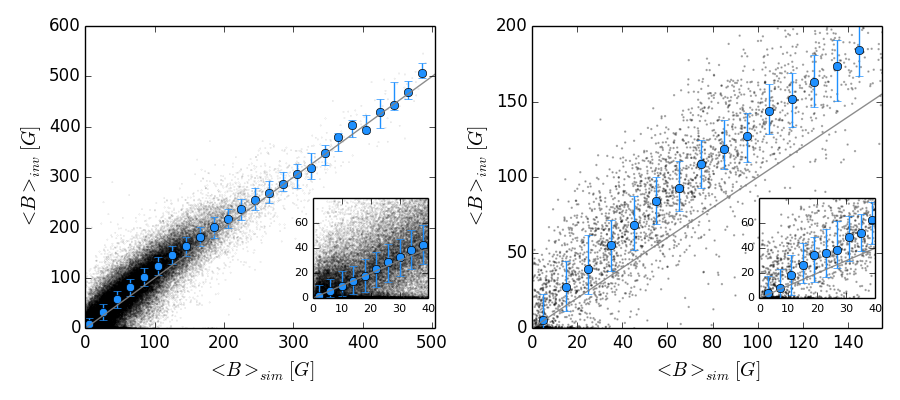}
\caption{Scatter plots of the magnetic field strength averaged over the formation region in the simulation versus the one inferred from the inversion of the synthetic Stokes I profiles. Left panel displays the case of the full resolution simulation test, while right panel corresponds to the degraded simulations test. In this last case, the quantity retrieved from inversions is the field strength that reproduces a particular averaged Stokes I profile. But since radiative transfer is not linear, this quantity has no counterpart in the degraded simulation. We have decided to compare the inferred field with the average magnetic field strength within the simulations larger pixel. The size of the bins used to represent the statistics of the scattered points are 20 and 10 G for left and right panel, respectively, and 4 G for both inset windows. The dots represent the 50th percentile in each bin, and the vertical lines englobe the 25th and 75th percentiles. The grey line displays the diagonal of the plot. Inset windows show a zoom for the weakest fields. \label{fig:two}}
\end{figure}

\subsection{Spatially degraded, noisy case}

But the challenge is to recover the average magnetic field strength when the fields are unresolved at the spatial resolution of the present data. We resample the simulations to the GRIS pixel size (about 100 km) to mimic the loss of information due to the finite spatial resolution of present, state-of-the-art, spectro-polarimetric data. We also add Gaussian noise at a level of the real observations. Taking into account that velocities are mixed when resampling, we allow for variations with depth of the microturbulence parameter. Figure 3 shows that the retrieval of magnetic field strength is greatly affected by the binning when inverting simultaneously the four Stokes parameters. As expected, since the actual pixel size is larger than the spatial variation of the magnetic field, there is a partial or complete loss of information because the Stokes parameters are signed. Nicely, good results are obtained when inverting only the intensity. Though the fields are a bit more spread and slightly stronger than in the simulation, the agreement is evident. Note that, in this case, the comparison of the inferred fields and those from the simulation is not straightforward. The lack of spatial resolution mixes the Stokes parameters, and the field we infer is the one that reproduces such averaged Stokes profile. But there is not an equivalent quantity in the degraded simulation. We therefore compare the inferred magnetic field with the average field in the resampled pixel of the simulations. Right panel of Fig. 2 shows the scatter plot of such values where we can see that the inferred value is slightly larger than the average field of the simulation.

\begin{figure}[ht!]
\plotone{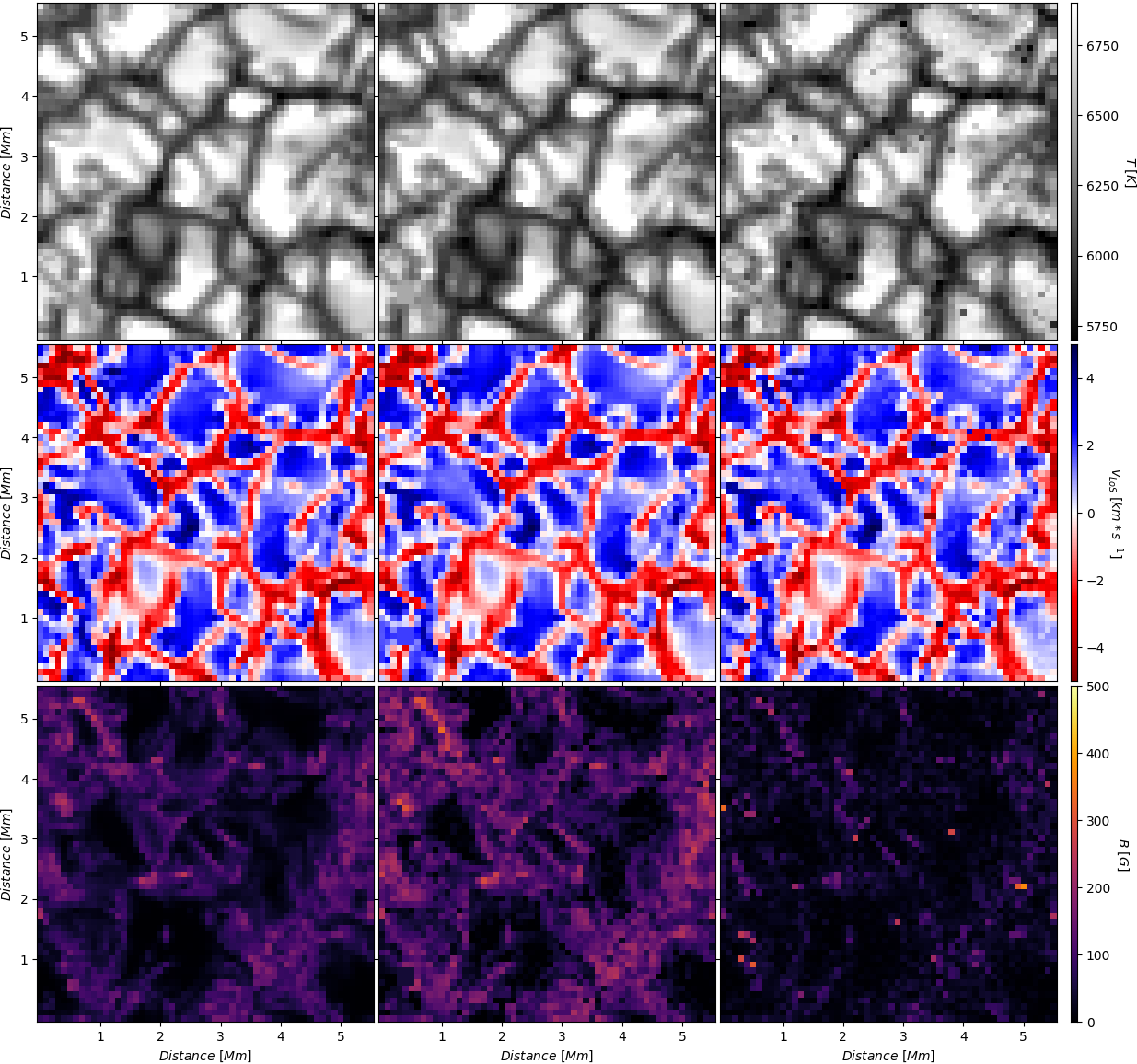}
\caption{Comparison between some parameters of the simulation (left column) averaged within the larger pixel to match the pixel size of GRIS observations, and the same parameters inferred from the inversion of the Stokes I profiles (middle column) and of the inversion of Stokes I, Q, U and V profiles simultaneously (right column). From top to bottom: temperature (at log($\tau_{500}$) = 0.0), line of sight velocity (at log($\tau_{500}$) = -0.2), and magnetic field strength averaged along the formation region of the spectral lines (from log($\tau_{500}$) = 0 to log($\tau_{500}$) = -1.2).\label{fig:three}}
\end{figure}

\section{Inversion of observed GRIS data} \label{sec:observations}

After validating our technique with numerical simulations, we infer the average magnetic field strength from the observations using only the intensity of many spectral lines. 
Observations come with some interpretation challenges since the Point Spread Function (PSF) of the telescope and the instrument, as well as that of the atmospheric seeing, couples the information of one pixel to the rest of the pixels in the field of view. The PSF typically has a core, that mixes the information of neighboring pixels. In ground-based telescopes, the width of this central core is dominated by the seeing conditions. The PSF has also extended tails, that account for an almost unpolarized profile. In order to remove the effect of the PSF in each pixel, ideally, we have to deconvolve the data \citep{valentin92,vannoort05,beck11,vannoort12,basilio13}. However, in general, the PSF of ground-based telescopes is unknown (except for some particular observational setups designed for that purpose \citep{vannoort17}) and changes with time due to seeing variations. Also, the deconvolution is a very ill-posed problem and, for quiet regions, many artifacts appear in the deconvolved data. Finally, the deconvolution increases the uncertainty, hence, it is not a very suitable technique for low-signal data, such as quiet areas. The deconvolution of observational data has only been performed successfully in satellite data, where the PSF is measured in laboratory before the flight and the time-dependence is negligible \citep{vannoort12,basilio13}. We use an approximation called local stray light \citep{orozco07}, and assume that the Stokes I spectrum in one pixel is contaminated mainly by the neighboring pixels. This contamination is computed using a 20x20 pixel box surrounding each pixel, and account for the 77\% of the resolution element. The way we reach these numbers is the following. We convolve the numerical simulations with the estimation of the GREGOR+GRIS+atmosphere PSF and then invert the synthetic profiles with different percentages of this local stray-light to match the physical conditions of the simulations. The 77\% of contamination is consistent with previous estimations of the stray light for GRIS data of about 80\%, obtained so that the continuum contrast of numerical simulations match the one of observations \citep{lagg16}. Using the stray-light approximation with a 77\% of filling factor, we find a mean magnetic field value (averaged over the field of view and in depth) for the convolved, rebinned simulations of 64 G. The value in the original, full resolution simulation was 69 G, hence proving the validity of our approach. As a robustness test, we have modified the stray light percentage values and checked that for values close to 77\% the inferred line of sight velocities averaged over the field of view for both in granules and in intergranules increase with depth, which is expected from theoretical considerations on convective motions. We therefore, modeled the intensity profiles as the result of a magnetized atmosphere that fills the 23\% of the pixel. Figure 4 displays the best fit of a randomly chosen observed profile. The goodness of the fit is evident. The model is flexible enough to fit the profiles of the 15 lines simultaneously, but, importantly, there is no overfitting to the data.

\begin{figure}[ht!]
\plotone{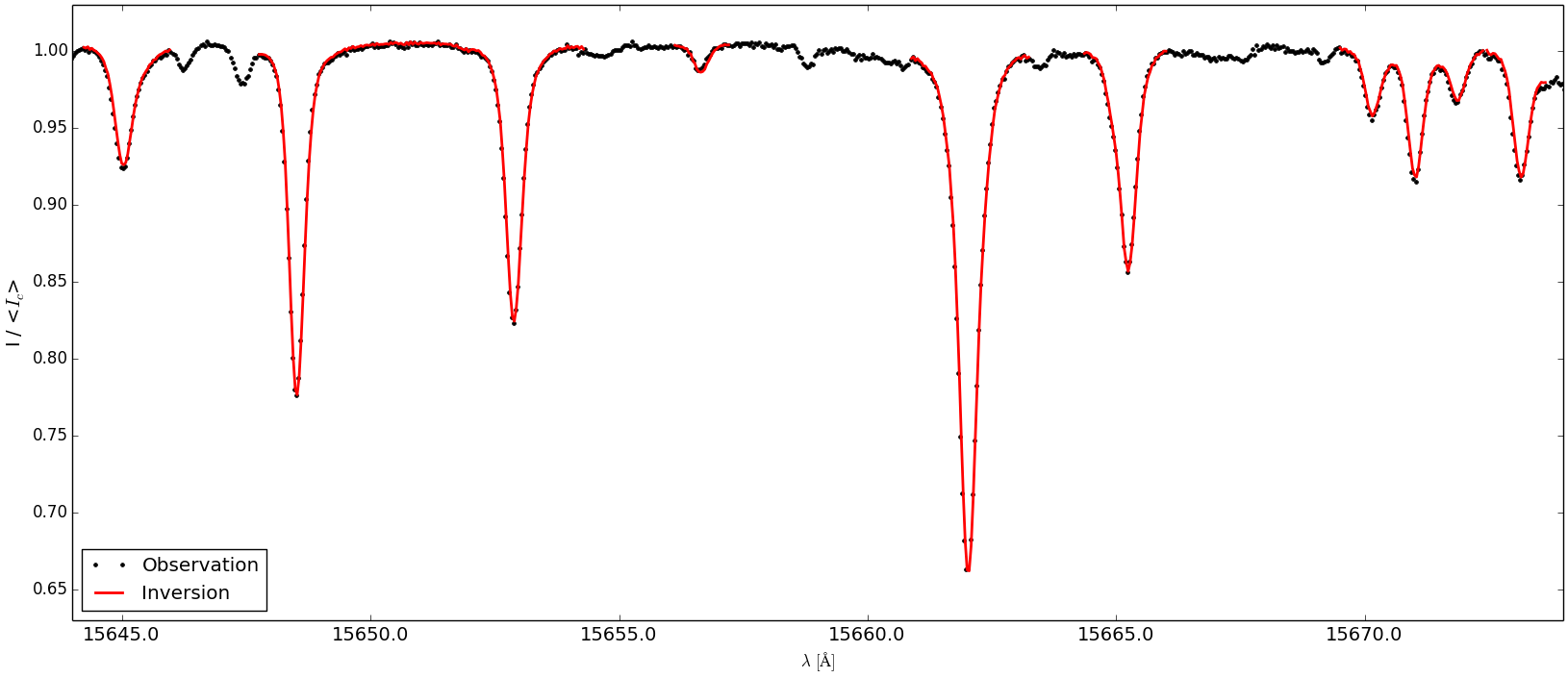}
\caption{Example of an inversion of an observational intensity profile. The observed GRIS spectrum is represented by black dots and the red curve is the best fit.\label{fig:four}}
\end{figure}

 Figure 5 shows the spatial distribution of the average field strength of the hidden quiet magnetism. There is an evident organization across the surface. Granular areas are almost devoid of magnetic fields, yet fields are concentrated in the intergranular lanes, with many patches of strong hecto- and kilo-Gauss fields. Convective motions are very likely behind this structure. Magnetic field lines are dragged by the laminar, more horizontal flow of granules, likely forming horizontal (hence low-lying) structures above them. These fields are transported to intergranular lanes where they are buffeted continuously, favoring the coalescence and amplification of the magnetic field to values higher than the equipartition. These fields tend to be more vertical because of floatability \citep{zwaan87}, allowing them to reach higher layers. The field-of-view averaged magnetization of granules and intergranules is 16 G and 76 G, respectively. In turn, this implies a global magnetization of 46 G.

\begin{figure}[ht!]
\plotone{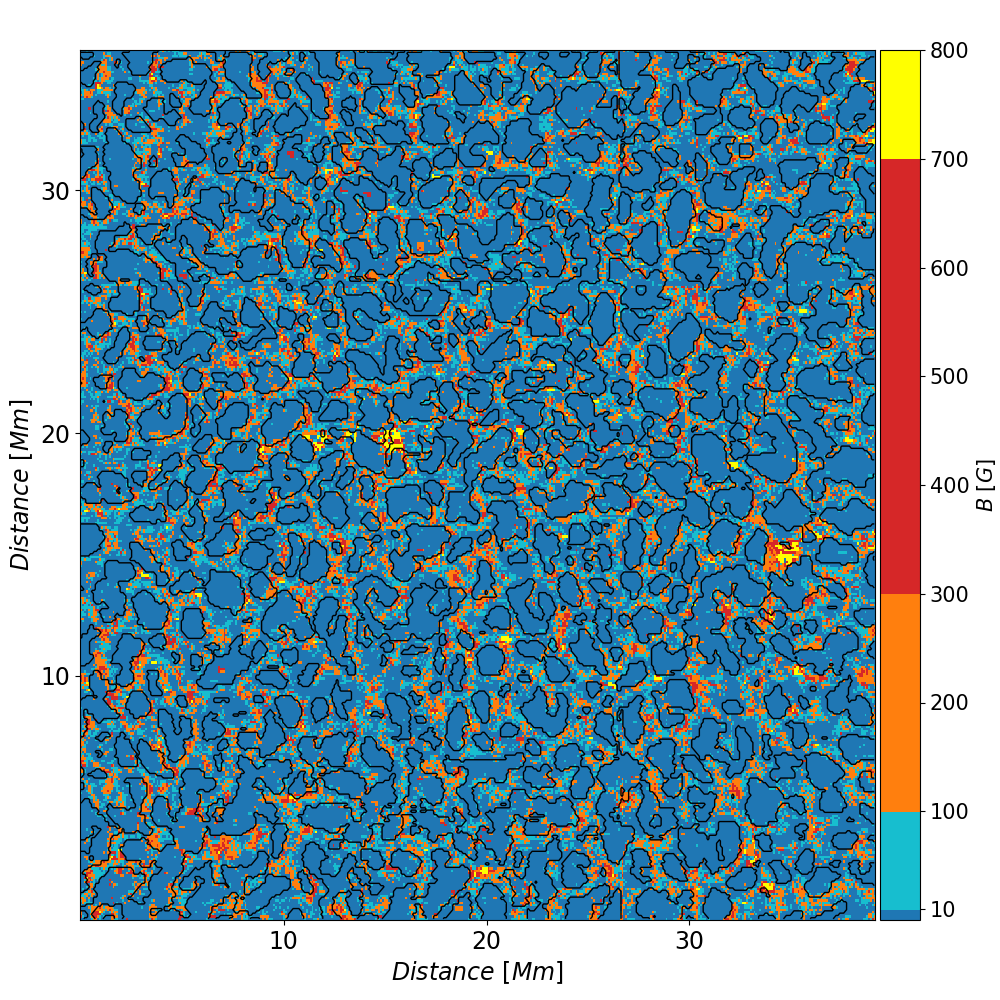}
\caption{Average magnetic field of the quiet Sun as inferred from the observational data. Black contours display the frontier between granules and intergranules (i.e., $I_{c}$ = 1.0, where $I_{c}$ is the continuum intensity). The magnetic field strength color bar has been clipped at 800 G.\label{fig:five}}
\end{figure}

\section{Conclusions and Discussion}

 We have relied on the Zeeman effect and performed intensity-only inversions to reveal the spatial distribution of the average strength of the hidden quiet Sun magnetism. Reaching such an important goal has required the inversion of many spectral lines simultaneously and the precise determination of atomic parameters \citep{trelles21}. The hidden field has a spatial variation clearly correlated with the convection phenomena of granulation, with granules almost devoided of fields and intergranules populated by much stronger fields. On average, the quiet Sun magnetism has a strength of 46 G. Assuming an Alfvén velocity of about 1-10 $km\ s^{-1}$, the stored energy in the quiet Sun, if transported upwards, is enough to compensate for radiative losses in the chromosphere (about 107 $erg\ cm^{-3}\ s^{-1}$). 
 
 Our empirical results can be contrasted with the most recent theoretical predictions, where a magnetization of about 100 G is needed to account for the observed (spatially unresolved) scattering signals in the Sr I line \citep{trujillo04,delpino18}. This line is, however, formed higher in the photosphere, even at the disk center, where the near infrared lines have very low, if any, sensitivity to the magnetic field. If we trust Hanle measurements and predictions, an increase of the average field strength with height in the atmosphere is required for Zeeman-based and Hanle-based results to be compatible \citep{trujillo04,delpino18}. Further work, as the study of the quiet Sun at different heliocentric angles to perform a tomography of the field with height, or the precise determination of collisions implicated in Hanle inference \citep[see][]{delpino18}, is required to solve this controversy.
 
 Such small-scale magnetism can never be observed in a star than the Sun, which strengthens the importance of our result. Nowadays, the most recent estimations of the hidden fields, in particular in young Suns, still rely on simpler techniques based on the measurement of the width and depth of some spectral lines \citep{kochukhov20}. Nicely, our technique can be extended with some future work to the Stellar Physics field and used to infer the global magnetization of other stars than ours.

\begin{acknowledgements}
The authors are especially grateful to Dr. Carlos Allende Prieto, Dr. Andr\'es Asensio Ramos, Prof. Manuel Collados Vera, and Prof. Javier Trujillo Bueno for very interesting discussions, carefully reading the manuscript, and helping with their comments to strengthen the conclusions of our work. We acknowledge financial support from the Spanish Ministerio de Ciencia, Innovaci\'on y Universidades through projects PGC2018-102108-B-I00 and FEDER funds. JCTA acknowledges financial support by the Instituto de Astrof\'isica de Canarias through Astrof\'isicos Residentes fellowship. MJMG acknowledges financial support through the Ram\'on y Cajal fellowship. BRC acknowledges financial support from the Spanish Ministerio de Ciencia e Innovaci\'on through the project RTI-2018-096886-B-C53. The observations used in this study were taken with the GREGOR telescope, located at Teide observatory (Spain). The 1.5-meter GREGOR solar telescope was built by a German consortium under the leadership of the Leibniz-Institute for Solar Physics (KIS) in Freiburg with the Leibniz Institute for Astrophysics Potsdam, the Institute for Astrophysics Göttingen, and the Max Planck Institute for Solar System Research in Göttingen as partners, and with contributions by the Instituto de Astrofísica de Canarias and the Astronomical Institute of the Academy of Sciences of the Czech Republic. This paper made use of the IAC Supercomputing facility HTCondor (http://research.cs.wisc.edu/htcondor/), partly financed by the Ministry of Economy and Competitiveness with FEDER funds, code IACA13-3E-2493.
\end{acknowledgements}

\bibliography{apj}

\end{document}